\documentstyle[11pt,newpasp,twoside]{article}

\begin{document}

\title{Theory of thermal and ionization effects in colliding winds of 
       WR+O binaries}
\author{Doris Folini\altaffilmark{1}}
\affil{Seminar of Applied Mathematics, ETH Zentrum,
       8092 Z\"{u}rich, Switzerland}
\author{Rolf Walder}
\affil{Institute of Astronomy, ETH Zentrum, 
       8092 Z\"{u}rich, Switzerland}

\altaffiltext{1}{Office currently at: Institute of Astronomy, 
                ETH Zentrum, 8092 Z\"{u}rich, Switzerland}

\begin{abstract}
The colliding winds interaction zone in WR+O binaries is a highly
complex environment. In this review we summarize the progress made
towards its theoretical understanding during the last years.  We
review the effect of different physical processes on the interaction
zone, among them geometry and orbital motion, radiative forces,
thermal conduction, instabilities and turbulence, ionizing radiation,
dust formation, clumped winds, magnetic fields, and particle
acceleration. Implications with regard to observations are
discussed. Subsequently, we proceed to the important question of
mutual interaction amongst these processes. Because of the wealth of
physical processes involved, numerical simulations are usually
mandatory.  Finally, we turn to the combined role these processes play
for the thermal and ionization properties of the colliding winds
interaction zone in WR+O binaries.
\end{abstract}
\section{Why bother? And how?}
Theory and observations both strongly suggest that colliding winds do
exist in WR+O binaries. This means that for a more complete
understanding of the physics of such binaries the collision zone must
be taken into account. Also, the presence of a collision zone in wide
binaries may allow to learn more about shock physics by comparing
theory with direct observations of the collision zone. Finally, the
presence of the collision zone is likely to contaminate many
observational quantities. Subsequently derived system parameters may
be contaminated as well. Seen in a more positive light, such
contamination could perhaps be used to detect previously unknown
binaries.

There exist various theoretical predictions for the influence of the
collision zone on observations, including enhanced X-ray emission
(Stevens 1992; Myasnikov \& Zhekov 1993; Pittard \& Stevens 1997;
Walder, Folini, \& Motamen 1999) and thermal radio emission (Stevens
1995), variability of line profiles in the UV (Shore \& Brown 1988;
Stevens 1993; Luehrs 1997), optical (Rauw, Vreux, \& Bohannan 1999),
and IR (Stevens 1999), the heating of the O-star photosphere (Gies,
Bagnuolo Jr., \& Penny 1997), and dust formation (Usov 1991).

The different models on which these predictions are based may be
divided into two groups. Starting from basic physics (e.g. Euler
equations) the interaction zone is modeled and some more or less
general predictions are made. The emphasis here lies on the
development and understanding of a consistent physical picture of the
interaction zone. In the following, we shall call them `type 1'
models. Instead, one may start from an observational feature and make
some assumptions about the interaction zone (e.g. geometrical shape,
ionization state, density) which are then tuned until the modeled
emission matches the observations. Here the point is to derive the
physical parameters of the interaction zone in accordance with
observations but basically without regard for what physics may be
responsible for the value of these parameters. Subsequently, we shall
call them `type 2' models. Both approaches have their merits and
drawbacks, and increased mutual exchange between them would seem
advantageous for both sides. This review, however, will mostly
concentrate on `type 1' models.
\section{Physical mechanisms at work: a visit to the zoo}
A variety of physical processes are of importance in colliding wind
WR+O binaries. In order to achieve a complete picture of the
situation, `type 1' models should include them all, an aim we are
still far away from today. In the following, some of the most relevant
processes are listed, along with a brief outline of their physical
implications and the state of the art with respect to their modeling.
\subsection{Geometry, orbital motion}
Obviously, reality takes place in three space dimensions. And some
observational features seem to require 3D models for their
explanation, e.g. the asymmetric X-ray light curve of $\gamma$ Velorum
(Willis, Schild, \& Stevens 1995) or the dust spiral in WR 104
observed by Tuthill, Monnier, \& Danchi (1999) which apparently even
rotates according to their observations. On the modeling side, there
are basically two approaches to 3D: by analytical means or through
numerical simulations.

Analytical approaches to 3D are often limited to approximate
geometrical descriptions of the location of the interaction
zone. Tuthill et al. (1999) point out, for example, that the spiral
observed in WR 104 follows approximately a rather simple geometrical
path, an Archimedian spiral. More elaborate analytical models
including orbital motion to some degree and describing the shape,
surface density, and velocity of the colliding winds interaction zone
have been provided by Cant\'{o}, Raga, \& Wilkin (1996) and Chen,
Bandiera, \& Wang (1996).  The analytical model of Usov (1995) goes
even further and allows, for example, predictions for X-ray emission
and particle acceleration.

The strength of such analytical models is that they yield precise
results within the frame of their assumptions, there are no artifacts,
and that they can be quickly evaluated for a particular set of
parameters. Their drawback is that they can take into account only a
few physical processes at a time and that they usually have to neglect
time dependence. While on geometrical grounds, for example, they
support the idea that the dust observed in WR 104 is related to the
interaction zone, they give no clue to why there should be dust at
all.

It is this latter kind of question where numerical models are needed
and where for some issues nothing less than 3D hydrodynamical
simulations will do. However, while obviously being closer to reality,
3D models are expensive in terms of CPU and memory requirements and
complicated with regard to data management, visualization, and
analysis of the simulations.  Consequently, they should be invoked
only where required. This point is also reflected in the fact that so
far only a handful of 3D hydrodynamical models exist for WR+O
binaries. First 3D simulations for three different WR+O binaries,
revealing the spiral shape of the interaction zone, were presented by
Walder (1995).  Based on 3D simulations of $\gamma$ Velorum Walder,
Folini, \& Motamen (1999) were able to obtain an asymmetric X-ray
light curve similar to the observed one. Pittard (1999) presented the
first 3D simulations including radiative forces acting on the winds in
the frame of CAK model. While much physics is still missing, these
simulations are on the edge of what is feasible today and they provide
a wealth of new insight.
\subsection{Radiative forces}
In close WR+O and O+O binaries the stellar radiation fields are strong
enough to affect the dynamics of the colliding winds. For comparable
stellar radii of the components, as in O+O binaries, radiative
inhibition can occur, as described by Stevens \& Pollock (1994). Here
the radiation field of one star can inhibit the acceleration of
the wind from the companion star. If the two stellar components have
largely different radii, as is probably the case in WR+O binaries,
Owocki \& Gayley (1995) demonstrated that radiative inhibition of the
WR-wind by the O-star radiation field is not efficient. Instead, 
radiative braking of the fully accelerated WR-wind occurs as it
approaches the O-star. Whether this braking of a highly supersonic
flow can be achieved without the generation of shocks is not yet clear.

In both cases, the stellar winds finally do not collide at the
terminal velocities of single star winds, but at lower
velocities. Consequently, the X-ray emission is softer and the total
X-ray flux is probably diminished as well. Owocki \& Gayley (1995)
also make the point that the opening angle of the collision zone can
be considerably increased due to radiative braking and that radiative
braking can prevent photospheric collision.

In both works CAK theory is applied to compute the radiative forces,
and within this frame both mechanisms suffer from the same main
difficulty: as the ionization state, temperature, and composition of
the matter is not exactly known, its response to the stellar radiation
fields and, therefore, the CAK coefficients are not well known
either. Despite these uncertainties Gayley, Owocki, \& Cranmer (1997)
have estimated that radiative braking should probably be taken into
account in a variety of close WR+O systems.
\subsection{Thermal conduction}
So far, thermal conduction by electrons and ions has mostly been
neglected when modeling WR+O colliding wind binaries. However, because
of its strong temperature dependence ($\partial_{t}T \propto
\nabla(T^{5/2} \nabla T)$) and the high post shock temperatures
reached in such systems (up to 10$^{8}$ K if terminal wind velocities
are reached) thermal conduction is likely to play an important role in
the physical description of the collision zone.

Quantitatively, not much is known about the influence of heat
conduction in WR+O binaries.  Myasnikov \& Zhekov (1998) have
performed 2D numerical simulations using a one temperature model. They
neglect radiative cooling and saturation effects, which in particular
also means that the entire interaction zone stays hot in their
simulations. Their results confirm most expectations: Pre-heating
zones, also known as thermal precursors, form upstream of each shock
whose temperatures and extensions depend significantly on the flow
parameters and the efficiency of thermal conduction. Meanwhile, the
temperature of the interaction zone decreases by up to an order of
magnitude compared to its adiabatic value. To preserve pressure
balance, its density increases by the same amount. The shocks become
isothermal. Myasnikov \& Zhekov (1998) also note that in the frame of
their model the growth of KH instabilities is reduced by strong heat
conduction.

The 1D simulations of Motamen, Walder, \& Folini (1999) show a drastic
change of the picture if radiative cooling is included.  A cold, high
density region can now form in the interaction zone.  The combination
of reduced post shock temperature and higher post shock density, both
due to thermal conduction, results in enhanced radiative cooling and
narrower cooling layers. For WR+O binaries this means that efficient
radiative cooling may already set in close to the center of the
system. Previous adiabatic cooling of the shocked matter to reach
temperatures where strong radiative cooling finally is possible may
become obsolete. In summary, the system is likely to become more
radiative under the combined influence of thermal conduction and
radiative cooling than if thermal conduction were absent.  For the
case of wind-blown bubbles, 1D simulations by Zhekov \& Myasnikov
(1998) suggest that the combined influence of radiative cooling and
thermal conduction can also cause the formation of additional multiple
shocks.

So far, there exist no 2D simulations for WR+O binaries including both
radiative cooling and thermal conduction. However, for other situations
such simulations have been performed, for example by Comer\'{o}n \&
Kaper (1998) for runaway OB stars.

In reality, however, thermal conduction is likely to be even more
complicated. Only two papers shall be mentioned here which illustrate
this. One is by Balbus (1986), where the emphasis lies on the effect
of magnetic fields and, in particular, on the time scales associated
with magnetized conduction fronts. The second is by Borkowski, Shull,
\& McKee (1989) who show that in many cases one temperature models
will not do. Also, they emphasize that the chemical composition and
the ionization state can affect the conductive heat flux and, in
particular, its saturation. They show that under certain circumstances
thermal conduction may be able to reduce the peak temperature only by
a factor of three, while for other conditions a reduction by a factor
of ten is possible.

With regard to observations, thermal conduction will certainly lead to
softer X-ray emission. One may speculate that the X-ray emission could
be enhanced as well. Due to higher compression and lower temperatures,
the shocked matter may cool radiatively before significantly cooling
adiabatically when moving out of the center of the system.
\subsection{Instabilities and turbulence} 
A wealth of analytical estimates and numerical simulations suggest
that the interaction zone in colliding wind WR+O binaries is 
unstable, especially when strong radiative cooling occurs. The
interaction zone as a whole gets bent and is possibly torn apart as
becomes apparent, for example, in the work of Stevens, Blondin, \&
Pollock (1992). More recent numerical simulations also suggest the
cold interior of the interaction zone to be in supersonically
turbulent motion.

A variety of papers are concerned with the physical nature of
different kinds of instabilities. The scope covered by these works
ranges from the classical Rayleigh-Taylor, Kelvin-Helmholtz, and
Richtmyer-Meshkov instabilities, which act on interfaces separating
two different physical states (for example the two winds), over
various thin shell instabilities, which act on the entire thin layer
of high density, cold matter (e.g. Dgani, Walder, \& Nussbaumer 1993;
Vishniac 1994), to the thermal instability related to radiative
cooling (e.g. Walder \& Folini 1995). A recent review can be found in
Walder \& Folini (1998).

Which of the suggested instabilities is important for a certain
astrophysical object is often not clear. Also, instabilities usually
will not occur isolated but will interact with each other, making it
often pointless to speak of one particular instability.  Finally, in
WR+O binaries advection out of the system center will be superimposed
on all instabilities (Belov, \& Myasnikov 1999; Ruderman 2000).

The resulting bending of the thin, cold, high density interaction zone
probably also affects the interior dynamics of this thin sheet. In a
planar, high resolution study Folini \& Walder (2000) have focused on
the interior structure of the cold part of a colliding winds
interaction zone. They find that the cold part of the interaction zone
to be subject to driven, supersonic turbulence. The matter
distribution within the turbulent interaction zone consist of
overcompressed, high density knots and filaments, separated by large
voids. The mean density of the interaction zone is considerably
reduced compared to the density required to balance the ram pressure
of the incoming flows by thermal pressure alone. Its surface becomes
billowy due to the turbulent motion inside.

The unstable behaviour of the cold part of the interaction zone should
have observable consequences. Spectral lines originating from it
should show clearly stronger than thermal line broadening because of
turbulent motion. The total extent of the hot post shock zones will be
affected by the combined influence of bending and the thermal
instability.  This may lead to some X-ray variability, as suggested by
Pittard \& Stevens (1997) for the case of O+O binaries.

Neglecting for a moment the review type character of this article,
some speculations may be added. On the basis of observed line profile
variations, Luehrs (1997) derived the angular extension of the
collision zone in HD 152270. The value he found seems extraordinarily
large for a cold, high density interaction zone. Could strong global
bending and twisting of a slim interaction zone make it appear to be
much more extended? The second speculation is related to the
overcompressed knots and filaments observed in numerical simulations
of radiatively cooling, unstable collision zones. Could dust formation
in close WR+O binaries be linked to such knots? (See also
Section~\ref{sec:dust})

Having argued in favor of an unstable interaction zone in colliding
wind WR+O binaries a word of caution seems advisable here as
well. Analytical results generally are not directly applicable because
their rather restrictive assumptions are usually not
fulfilled. Numerical simulations, on the other hand, are more flexible
in that respect but it may be difficult to rule out numerical
artifacts. Also, instabilities so far have mostly been studied in the
frame of rather simple physical models. How additional physical
processes will influence the stability properties of the interaction
zone has yet to be investigated.

With regard to possible numerical artifacts the recent publication of
Myasnikov, Zhekov, \& Belov (1998) must be mentioned. They argue that
if not purely numerical in origin anyway, the instabilities observed
in numerical simulations of colliding wind binaries at least greatly
depend on the applied cooling limit. Their findings certainly require
further attention.

Despite these objections, we are firmly convinced that the colliding
winds interaction zone in WR+O binaries is unstable. If efficient
radiative cooling takes place the interaction zone is most likely
subject to strong bending and is possibly torn apart in some
locations. We are, however, not sure how violent this instability is
and what its exact physical cause is. The cold interior of the
interaction zone is probably subject to supersonic turbulence. How
exactly this turbulence is driven is not yet clear. Also, the
statistical properties of this part of the interaction zone, for
example its mean density, have barely been investigated and then only
in 2D. This unstable behaviour must cause observable traces.
\subsection{Ionizing radiation} 
There are three sources of ionizing radiation in WR+O binaries: the
two stars themselves and the shock heated interaction zone. For the
temperature and ionization state of the cold matter within and around
the colliding winds interaction zone this radiation is crucial. So
far, only a few studies exist, each of which deals with a separate
aspect of the problem. Gies, Bagnuolo Jr., \& Penny (1997) find that
the X-ray radiation emitted by the collision zone in close binaries is
capable of significantly heating the O-star photosphere, thereby
changing observational quantities used to derive the stellar
parameters. Aleksandrova \& Bychkov (1998) considered the same
radiation source but investigated its influence on the pre-shock
material. Investigating wind velocities in the range between 4000 km/s
and 15000 km/s they found that the X-rays from the collision zone are
capable of ionizing iron nearly completely {\it before} it gets
shocked. While these velocities are clearly above those encountered in
WR+O binaries the pre-ionizing effect as such is likely to be present
also in these systems. Consequently, emission from highly ionized ions
may not solely originate from the shock heated zones themselves, which
should be taken into account when using highly ionized elements for
diagnostics of the collision zone. Also the temperature of the
pre-shock matter is likely to be affected. First attempts to estimate
the effect of the stellar radiation field on the cold part of the
interaction zone have been made by Rauw et al. (1999) and
by Folini \& Walder (1999), the latter using 3D optically thick NLTE
radiative transfer. Although their results are preliminary, the latter
authors find that for their toy example of $\gamma$ Vel optically
thick effects become important within the cold, high density part of
the collision zone.
\subsection{Dust formation}
\label{sec:dust}
Observations clearly prove the permanent or episodical dust formation
in certain WC+O binaries. The most spectacular example is probably the
dust spiral of WR 104 observed by Tuthill et al. (1999).  Recent
observational summaries can be found in Williams (1997) and Williams
(1999). The observation of dust in such systems is puzzling as
conditions there (high temperature, strong UV radiation) seem not
especially suited for dust formation.

Theoretically, dust formation in WR+O binaries is not understood so
far. Usov (1991) has published density estimates for a homogeneous
collision zone. The 2D simulations of V444 first presented by Walder
\& Folini (1995) have been carried on in the mean time, showing
overcompressions of up to a factor of ten in a supersonically
turbulent interaction zone, compared to a homogeneous one. In this
particular simulation densities of up to about 10$^{13}$ cm$^{-3}$ are
observed out to a distance comparable to the separation of the two
stars. Considerably more publications exist on dust nucleation and
grain growth under laboratory conditions and in WC winds. Cherchneff
\& Tielens (1995) and Cherchneff (1997), for example, focused on dust
nucleation. In particular, they found that high densities, possibly up
to 10$^{12}$cm$^{-3}$, are required for dust nucleation to take place
while the nucleation process is nearly independent of temperature
between 1000 K and 4000 K. Although these temperatures seem small for
the colliding winds interaction zone in WR+O binaries they might not
be out of reach if the densities are high enough. (See also
Section~\ref{sec:open}) Leaving the question of nucleation aside,
Zubko (1992), Zubko, Marchenko, \& Nugis (1992), and Zubko (1998)
carried out theoretical studies of grain growth via collisions of
charged grains with carbon ions in WC winds. A main conclusion from
their work is that dust grains may grow even in a highly ionized
standard WC atmosphere, provided the condensation nuclei are created
somehow.
\subsection{Clumped winds, magnetic fields, particle acceleration}
The influence of several other physical processes on the collision
zone is even less investigated than of those outlined in the previous
sections. Three of them, clumped winds, magnetic fields, and particle
acceleration, shall be briefly touched in the following.

Evidence is growing that the winds of both, WR- and O-stars are indeed
clumped rather than smooth. However, the size, compactness and
distribution of the clumps is still under debate. Is a clumped wind
more like a few massive clumps in a homogeneous flow? Or is more
appropriate to talk about the flow in terms of an ensemble of
different blobs?  As pointed out for example by L\'{e}pin (1995), the
effect of clumped winds on the interaction zone will depend on which
of the two scenarios applies. A fast, compact, high density clump may
pass through the entire interaction zone with basically no interaction
at all (see e.g. Cherepashchuk 1990).  A less dense, not too fast
clump, on the other hand, may finally get dissolved in the interaction
zone, thereby possibly affecting its stability and emission. But also
the theoretical treatment may be different, depending on the nature of
the clumped winds. Can they be treated statistically using some mean
properties or is it important to treat clumps as 'individuals'?

Concerning magnetic fields, the observation of non-thermal emission
suggests the presence of magnetic fields in at least some WR+O
binaries. However, their strength and orientation have yet to be
determined. On the theoretical side, only a few papers exist on
magnetic fields in colliding wind WR+O binaries. Eichler \& Usov
(1993) and Jardine, Allen, \& Pollock (1996) present studies on
particle acceleration and related synchrotron emission in the
interaction zone in WR+O binaries. Zhekov, Myasnikov, \& Barsky (1999)
focus on the magnetic field distribution, assuming for the stellar
wind magnetic field a simplified Parker model. Depending on their
strength, magnetic fields are likely to affect several other physical
processes directly or indirectly as well.  (See also
Section~\ref{sec:open})
\section{Open the fences: mutual interactions}
\label{sec:open}
The different physical processes addressed in the previous section
obviously do not occur as isolated processes. Some processes influence
others and vice versa. A few examples of this we have already briefly
encountered above. However, there exist hardly any models including
more than one or two different physical processes at the same time.
It just would be too costly up to now to include more physics in the
numerical models. Also, it may be wiser anyway to first improve our
understanding of simpler situations before turning to more complicated
ones. Nevertheless, one should bear in mind that most likely there
will be considerable interaction amongst different physical
processes. The remainder of this section is devoted to speculations,
rather than results, on a few such interactions. This section may,
therefore, be considered to go beyond the frame of a review. However,
we deem it necessary to address these questions as they are crucial
for the physical understanding of the interaction zone.

Consider first thermal conduction and radiative braking. Especially
for close binaries, they are both crucial for the post shock
temperature as well as for the location of the interaction zone and
its opening angle. The interaction of the pre-shock matter with the
radiation field, crucial for radiative braking, will certainly be
affected by the increase in temperature due to thermal conduction. It
seems plausible to assume that this interaction will be reduced if the
matter temperature deviates strongly from the radiation temperature.
Let us first start from a radiative braking model. If we now add
thermal conduction the pre-shock wind will be heated. According to our
assumption this means less interaction between the wind and the
radiation field and therefore less radiative braking. The post shock
temperature will rise and the pre-heating zones will become even
larger. 'Positive back coupling' occurs. Now let us start with a
thermal conduction model. Adding radiative braking will slow down the
pre-shock wind, the post shock temperature will drop, and the
pre-heating zones will become smaller and cooler. Again according to
our assumption, radiative braking will become more efficient and the
post shock temperature will drop. 'Positive back coupling' again. Both
scenarios are speculations. But as each of the two processes alone
already has powerful impact on the physics of the interaction zone a
common investigation of the two would be highly desirable.

Here it may be added that a radiative precursor instead of a thermal
one could have similar effects. A radiative precursor is likely to
exist around the wind collision zone in WR+O binaries because of the
X-ray and UV emission of the high temperature post shock zones.  Its
effect would be to heat and ionize the pre-shock matter.  However, at
least for the case of colliding winds in WR+O binaries radiative
precursors on their own are even less investigated up to now than
thermal ones.

Another issue is the interplay of thermal conduction, radiative
cooling, and ionizing radiation. Together these processes essentially
determine how cold the matter can get in the cold, high density part
of the interaction zone. While attempts have been made to bring the
first two together the isolated problem of the influence of ionizing
radiation on the interaction zone has hardly been investigated so
far. Its influence is usually taken into account only in the form of
heating due to photoionization and then only in the form of a more or
less arbitrarily chosen cooling limit of the radiative loss
function. Just for arguments sake consider a compact, high density
clump in the interaction zone that is opaque to UV radiation. Its
outside would be bombarded by UV and X-ray photons, the surface of the
clump would be heated. Now, thermal conduction would tend to
distribute this energy, received on the surface, over the entire
volume of the clump. If the clump then were able to radiate this
energy at longer wavelengths, for which the clump were still
transparent, the clump may manage to remain cold. Such a mechanism
could possibly preserve the cold environment necessary for dust
formation.

So far, we have again neglected magnetic fields in this section.
Their presence, however, will have a direct influence on thermal
conduction or the stability and density of the cold part of the
interaction zone. The altered thermal conduction then in turn may
affect, for example, radiative braking or dust formation. So
indirectly magnetic fields may influence physical processes or
quantities, like radiative braking, which at first glance one may
believe to be unaffected.
\section{End of the visit: collecting the pieces}
Theoretical predictions for the thermal and ionization state of the
colliding winds interaction zone in WR+O star binaries require the
inclusion of a variety of physical processes. Which processes are
indeed important for which system is, however, often a difficult
question. A brief summary of where we stand today is attempted in the
following.

{\it How hot can it get?} Being one of the key questions it has been
examined quite carefully. The only process leading to a temperature
increase in the high temperature part of the interaction zone is shock
heating. The temperature reached depends on the relative velocity of
the colliding flows which, in turn, can be affected by radiative
braking. Thermal conduction, on the other hand, causes a decrease of
the peak temperature. This peak temperature may be higher than the
temperature we observe, depending on when radiative cooling becomes
important. If the density is too low or the peak temperature too high
the matter will undergo significant adiabatic cooling as it moves out
of the center of the system, before it starts to cool radiatively and
thus becomes observable.  The peak temperature may also affect the
maximum densities which can be reached within the cold part of the
collision zone. A lower peak temperature, due to thermal conduction
for example, allows for faster cooling, closer to the center of the
system, and consequently for higher densities of the cooled
matter. Presently, studies exist on the influence of each single
processes and on the combined influence of thermal conduction and
radiative cooling in 1D. The combined influence of all these processes
has, however, yet to be investigated, finally in 2D or 3D. The
question is certainly important with regard to X-ray
emission. Comparison with observation shows that for close binaries
the predicted X-ray emission is still too high and too hard, whereas
for wider systems theory and observation agree much better. The
combined influence of the above processes probably will help to bring
theory and observation closer together in this point.

{\it How cold can it get?} Another key question which, despite its
importance with regard to ionization states, compression, and dust
formation, especially for close binaries, has barely been attacked so
far. Basically, thermal conduction and photoionization tend to heat
the cold part of the interaction zone, whereas radiative cooling
reduces its temperature. From these processes, heating of the
interaction zone by photoionization is by far the least
investigated. Here the stellar radiation fields as well as the
radiation from the shock heated zones must be taken into account and
their interaction with the matter has to be determined.  At least for
some systems, this probably requires detailed multidimensional
radiative transfer computations.  The difficulty with radiative
cooling, crucial in this context, is that it strongly depends on the
temperature and ionization state of the matter. The cooling history,
and therefore time dependence, may also be important.  Finally, the
properties of the hot post shock zones, the source of thermal
electrons and ionizing photons, are not well known either. In the 
future, the question should be clarified whether detailed radiative 
transfer is indeed needed. Then, the heating of the cold part of the
interaction zone by photoionization should be investigated more 
quantitatively. 

{\it Towards observations.} Traces of the interaction zone seem to be
present in all spectral ranges. Observational predictions from what we
called `type 1' models in this review exist, to our knowledge, for
X-rays (many) and radio (one). As far as we know, there are no
predictions form `type 1' models for UV, optical, and IR. The reason
is that predictions for these latter spectral ranges depend essentially
on the cold part of the interaction zone which, as mentioned before,
is not yet well understood. All models reproducing line profile
variations in the UV, optical, or IR are of what we called `type 2'.
Starting from a more or less simple geometrical description of the
interaction zone these models then assume a certain ionization state
and level population. They are valuable as they show us what certain
parameters of the interaction zone should be like in order to
reproduce observations. However, these models themselves give no
physical explanation of why the parameters should have their
particular values.

{\it The 'grand unified model'.} Looking at `type 1' models, some
physical processes are included in one model but not in the other (for
example thermal conduction or radiative braking) and some physical
ingredients are barely considered at all so far (for example clumped
winds or magnetic fields). `Type 2' models may be improved by
considering other than just the most simple, analytical matter
distributions. It may also be worthwhile trying to find out how unique
a certain observed signature is. Is there only one way to reproduce it
or allow several different, equally plausible models for the same
observational feature? Enhanced combination of `type 1' and `type 2'
approaches could help to decide such questions. Although a 'grand
unified model' for WR+O colliding wind binaries will remain out of
reach for several years to come, considerable progress has been made
with regard to the modeling and understanding of single physical
processes such a model must comprise. This should also help us to
decide which physical processes are indeed needed in order to explain
a particular system.  Also in future modeling will be expensive and
models, therefore, should comprise only the essential physics.

\end{document}